\documentclass[reprint, amsmath, amssymb, aps, twocolumn]{revtex4}

\usepackage{graphicx}
\usepackage{dcolumn}
\usepackage[arrow, matrix,curve]{xy}
\usepackage{hyperref}

\newcommand{\AHI} {\A_{\mathrm{cosm}}}
\newcommand{\GEL} {\mathcal{G}}

\newcommand{\ovl}[1] {\overline{#1}}

\newcommand{\cp} {\circ}
\newcommand{\SU} {\mathrm{SU}(2)}

\newcommand{\VD} {\mathfrak{V}}

\newcommand{\DIDE} {\mathfrak{D}}

\newcommand{\act} {\theta}
\newcommand{\sact} {\Theta}

\newcommand{\CAP} {C_{\mathrm{AP}}(\RR)}
\newcommand{\AP} {\mathrm{AP}}

\newcommand{\A} {\mathcal{A}}

\newcommand{\X} {\Spec(\cC)}
\newcommand{\x} {\ovl{c}}
\newcommand{\ux} {\underline{c}}

\newcommand{\RR} {\mathbb{R}}
\newcommand{\RB} {\mathbb{R}_{\mathrm{Bohr}}}
\newcommand{\NN} {\mathbb{N}}

\newcommand{\CC} {\mathbb{C}}

\newcommand{\qcR} {\underline{\RR}}
\newcommand{\qR} {\ovl{\RR}}

\newcommand{\wpo} {\widehat{p}}
\newcommand{\gelhat}[1] {\widetilde{#1}}

\newcommand{\Transl} {\Lambda}
\newcommand{\Transw} {\boldsymbol{\Lambda}'}
\newcommand{\Transwq} {\boldsymbol{\Lambda}}

\newcommand{\BP} {\boldsymbol{+}}
\newcommand{\BM} {\boldsymbol{-}}

\newcommand{\Hil} {\mathcal{H}}

\newcommand{\NB} {0_{\mathrm{Bohr}}}
\newcommand{\muB} {\mu_{\mathrm{H}}}

\newcommand{\muR} {\mu_{\RR}}
\newcommand{\muRB} {\mu_{\mathrm{B}}}

\newcommand{\cC} {\mathfrak{C}}
\newcommand{\Spec} {\mathrm{Spec}}

\newcommand{\dd} {\mathrm{d}}
\newcommand{\ee} {\mathrm{e}}
\newcommand{\I} {\mathrm{i}}

\newcommand{\Borel} {\mathfrak{B}}

\newcommand{\itspace} {\vspace{0pt}}

\begin{document}

\title{Kinematical Uniqueness of Homogeneous Isotropic LQC}

\author{Jonathan Engle, Maximilian Hanusch}
 \email{jonathan.engle@fau.edu, hanuschm@fau.edu}
 
\affiliation{Physics Department, Florida Atlantic University, 777 Glades Road, FL 33431, USA}

\date{December 8, 2016}

\begin{abstract}

In a paper by Ashtekar and Campiglia, invariance under volume preserving residual diffeomorphisms 
has been used to single out the standard representation of the reduced holonomy-flux algebra in homogeneous loop quantum cosmology (LQC). In this paper, we use invariance under \textit{all}
residual diffeomorphisms to single out the standard kinematical Hilbert space of homogeneous
\textit{isotropic} LQC for \textit{both} the standard configuration space $\RB$, as well as for the Fleischhack one $\RR\sqcup \RB$. We first determine the scale invariant Radon measures on these spaces, and then show that the Haar measure on $\RB$ is the only such measure for which the momentum operator is hermitian w.r.t.\ the corresponding inner product.  In particular, the measure is forced to be identically zero on $\RR$ in the Fleischhack case, so that for both approaches, the standard kinematical LQC-Hilbert space is singled out. 
%
% Note that, because we impose only invariance of the measure under residual diffeomorphisms, and not invariance of the
% representation more generally, the issue of non-preservation of the quantum algebra by non-volume-preserving 
% diffeomorphisms is side-stepped. 
%
\end{abstract}

\maketitle

\section{Introduction}
\label{sec:intro}
Diffeomorphism invariance is the symmetry which guided Einstein to his general theory of relativity, and represents one of the 
deepest and most successful conceptual leaps in modern physics.  
As gauge, this symmetry entails that `space-time points' have no physical meaning, unless some feature of the geometry or matter uniquely distinguish it. 
Retention of this symmetry at the quantum level is the principle underlying the quantization of general relativity known as loop quantum gravity (LQG) \cite{thiemann2007, rovelli2004, BackLA}.

In order to both test the ideas in LQG, as well as to understand how LQG might resolve the classical Big Bang singularity and lead to potentially observable cosmological predictions, the simplified model known as loop quantum cosmology (LQC) was introduced
\cite{aps2006, MathStrucLQG, bojowald2008}. The framework of LQC is obtained by quantizing the homogeneous isotropic sector of classical gravity, coupled to matter, using methods similar to those of loop quantum gravity.  

Specifically, in LQG, one uses the Ashtekar-Barbero formulation of gravity \cite{barbero1995}, in terms of an $\SU$ connection $A^i_a$ and its conjugate densitized triad field $E^a_i$. The basic configuration variables directly promoted to operators in the quantum theory are the parallel transports of $A^i_a$ along piecewise analytic curves.  In order to construct a Hilbert space of quantum states which carries a representation of the algebra 
$\cC$ of these variables, one introduces what is called the \textit{Gel'fand spectrum} 
$\Spec(\cC)$ 
of $\cC$.  
This space represents a compactification of the classical configuration space $\mathcal{A}$ of smooth $\SU$ connections,
and thus is referred to as the \textit{quantum configuration space}, often denoted $\overline{\mathcal{A}}$.
The topology is such that the algebra of functions $\cC$ 
is precisely the algebra of continuous functions on $\overline{\mathcal{A}}$, 
whence it is much easier to construct Radon measures on 
this space. The condition that such measure be \textit{diffeomorphism invariant} uniquely selects the 
\textit{Ashtekar-Lewandowski} measure $\mu_{AL}$, and the Hilbert space of kinematical quantum states is
taken to be $L^2(\overline{\mathcal{A}}, \mu_{AL})$ \cite{OK}. Here, the operators that correspond to the algebra $\cC$, act by multiplication; and the conjugate momentum operators that correspond to fluxes of $E^a_i$, act by derivation.

In LQC, one starts with the homogeneous isotropic sector of Ashtekar-Barbero gravity, parameterized by a ``connection'' $c$
and conjugate momentum $p$ via $A^i_a = c \mathring{A}^i_a$ and $E^a_i = p \mathring{E}^a_i$ where 
$(\mathring{A}^i_a, \mathring{E}^a_i)$ are fixed. The quantization of this system then proceeds in a fashion identical to that for the full theory described
above, \textit{except} in two respects:
\begingroup 
\setlength{\leftmargini}{12pt}
\begin{enumerate}
\item 
	The algebra of elementary configuration variables $\cC$ is chosen to include only parallel transports along 
paths which are \textit{straight} (i.e.,  integral curves of the translational killing vector fields).  This leads to the
quantum configuration space $\RB$, the Bohr compactification of the real line.
\item 
	The measure on the quantum configuration space is selected by mathematical naturalness rather than through a physical symmetry principle.  Specifically: The Haar measure $\muB$ on $\RB$ is used, rather than selecting a measure through invariance under a physical symmetry.
\end{enumerate}
\endgroup
\noindent
Through the results of this paper, we show that both of these deviations from the LQG program can be rectified, and that the resulting kinematical Hilbert space representation is \textit{the same as before}.  That is, the above two deviations, originally made for simplicity, 
have no physical consequence. 
The only advantage gained by rectifying the above two deviations is a systematic definition of operators
corresponding to parallel transports along \textit{curved} paths on the standard Hilbert space of LQC -- a definition non-systematically guessed beforehand in the work \cite{engle2013}.
\clearpage

\subsubsection*{More precise summary of the results}
More precisely, in previous work \cite{CSL} Fleischhack has shown that by correcting Point 1. above, the quantum configuration space changes to $\RR\sqcup \RB$.  In this paper, we select the measure on $\RR\sqcup \RB$ by imposing 
\textit{residual diffeomorphism invariance}, similar to the strategy used in \cite{UniquOfKinLQC} for the non-isotropic case \footnote{This condition is physically more well-motivated than the condition used in \cite{UMLQC}.}. 
At the same time, we also select a measure on the standard configuration space $\RB$, using the same criterion. 

In the homogeneous isotropic 
context, the residual diffeomorphisms are simply the one-parameter group of \textit{dilations}, the action of which yields
the usual rescaling symmetry of the basic variables $(c,p) \mapsto (\lambda c, \, {\rm sgn}(\lambda) \lambda^{2} p)$ for 
$\lambda \in \RR_{\neq 0}$.  
In LQG, one imposes invariance of the measure under the action of diffeomorphisms on the configuration variable, which leads uniquely to the Ashtekar-Lewandowski measure. Likewise, in the present context, we impose invariance of the measure under the action of \textit{residual} diffeomorphisms on the configuration variable.  That is, we impose invariance of the measure (assumed to be Radon) under rescaling of $c$. Notice that, because we include negative rescalings in this invariance, we are also imposing invariance under \textit{orientation reversing} diffeomorphisms. 
Moreover, since the relevant spaces are compact, each Borel and hence Radon measure is necessarily finite, 
and so can be normalized without loss of generality.

It turns out, however, that imposing this invariance is not quite enough to yield uniqueness: It is necessary to use one other condition.  The condition we use is the existence of a momentum operator such that (1.) its commutator with the configuration operators mimics the corresponding Poisson brackets, and (2.) it is hermitian, as must be the case because it corresponds to a real quantity. This is then enough to yield uniqueness --- more precisely, to single out $\muB$ on both configuration spaces. Note that the only conditions used are very elementary conditions at the core of any quantization.

The result for the unique measure on $\RR \sqcup \RB$ is remarkable in its importance as it is equal to zero on the $\RR$ part, and equal $\muB$ on the $\RB$ part.  Thus, the resulting Hilbert space of kinematical states, namely the space of square integrable functions with respect to the determined measure (because it involves division by zero norm states) 
is naturally isomorphic to the space of square integrable functions on $\RB$ with respect to the Haar measure. That is, the resulting kinematical Hilbert space, and representation of the basic variables thereon, is exactly the standard one used in LQC up until now.

It is worthwhile to remark in more detail on the relation of these results to those of \cite{UniquOfKinLQC}. In the work \cite{UniquOfKinLQC}, the authors select not only the measure on the quantum configuration space using invariance under residual diffeomorphisms, but the entire representation of the quantum algebra using this condition.  Hence, an action of the residual diffeomorphisms on the quantum algebra is required, and for this reason they must restrict consideration to volume preserving diffeomorphisms, which have non-trivial action only in the non-isotropic case. In the present paper, 
by focusing only on invariance of the measure, the issue of a well-defined action on the quantum algebra is 
side-stepped, so that volume non-preserving diffeomorphisms can also be considered.  This is what allows the analysis
to be applied to the isotropic case, and hence in particular also to the Fleischhack framework.

\section{Configuration Spaces}
\label{sec:prel}
The quantum configuration space of LQG is given by the Gel'fand spectrum of the $C^*$-algebra generated by matrix entries of parallel transports $h_\gamma\colon \A\rightarrow \SU$ 
along embedded analytic curves in the base of the $\SU$-principal fibre bundle $P$ of interest. Here, $\A$ denotes the set of smooth connections on $P$; and since 
$\SU$ is compact, each such matrix entry is indeed a bounded function on $\A$. This ensures that the uniform closure of the $^*$-algebra generated by them exists and is $C^*$.

In the homogeneous isotropic case (LQC), only parallel transports w.r.t.\ homogeneous isotropic connections are quantized. Here, we have $P=\RR^3\times \SU$, and the quantum configuration space of interest is the Gel'fand spectrum of the $C^*$-algebra generated by the restrictions of $h_\gamma$ to $\AHI\cong \RR$. Traditionally, only parallel transports along straight curves are considered; yielding to the $C^*$-algebra $\CAP$ of almost periodic functions on $\RR$ with  generators $\chi_\lambda\colon c\mapsto \ee^{\I \lambda c}$. The spectrum of $\CAP$ is the compact abelian group $\RB$, whose continuous group structure is uniquely determined by
\begin{align*}
\begin{split}
	(\psi_1 \BP \psi_2)(\chi_\lambda)&:=\psi_1(\chi_\lambda)\cdot \psi_2(\chi_\lambda)\\
	\NB(\chi_\lambda)&:=1\\
	\BM\hspace{1pt} \psi(\chi_\lambda)&:=\overline{\psi(\chi_\lambda)}
\end{split}
\end{align*}
for all $\lambda\in \RR$, $\psi, \psi_1,\psi_2\in \RB$. Here, the bar denotes the complex conjugation. The homogeneous isotropic connections (parametrized by $\RR$) are densely embedded into $\RB$ via the homomorphism 
\begin{align}
\label{eq:iota}
\iota\colon \RR\rightarrow \RB,\quad c\mapsto [\varphi \mapsto \varphi(c)]
\end{align}
for $\varphi\in \CAP$; more precisely, meaning that $\iota$ is continuous and injective, that the closure of $\iota(\RR)$ is $\RB$, and that 
\begin{align}
\label{eq:homomorph}
	\iota(c+c')=\iota(c)\BP\iota(c')\qquad \forall\:c,c'\in \RR.
\end{align}
holds. 

In contrast to standard LQC where only parallel transports along straight curves are considered, in the Fleischhack approach, all embedded analytic curves are used to define the cosmological quantum configuration space. More precisely, this means that there the spectrum of the $C^*$-algebra generated by the restrictions $h_\gamma|_{\AHI\cong \RR}$ is considered, whereby $\gamma$ now runs over all embedded analytic curves in $\RR^3$. This $C^*$-algebra is given by the direct vector space sum $C_0(\RR)\oplus \CAP$ \cite{CSL}, and its spectrum is homeomorphic \cite{CSL} to the space  $\RR\sqcup \RB$ equipped with the compact topology generated by the sets   
\begin{align*}
  \begin{array}{rclcl}
    V & \!\!\!\sqcup\!\!\! & \emptyset 
    && \text{$V \subseteq \RR$ \hspace{1.4pt}open} \\
     K^c & \!\!\!\sqcup\!\!\! & \RB
    && \text{$K \subseteq \RR$ compact} \\[-1.3pt]
     \varphi^{-1}(U) & \!\!\!\sqcup\!\!\! & \mathcal{G}(\varphi)^{-1}(U) 
    && \text{$U \subseteq \mathbb{C}$ \hspace{1.4pt}open, $\varphi \in \CAP$}.
  \end{array}
\end{align*}
Here, $\GEL(\varphi)$ denotes the Gel'fand transform of $\varphi$ w.r.t.\ $\Spec(\CAP)$; and the respective homeomorphism 
\begin{align*}
\xi\colon \RR\sqcup \RB =:\qcR\rightarrow \qR:=\Spec(C_0(\RR)\oplus\CAP) 
\end{align*}
is given by \cite{CSL} 
\begin{equation}
  \label{eq:Ksiii}
  \xi(\ux) := 
  \begin{cases} 
    \hspace{34.9pt}\varphi\mapsto \varphi(\ux) &\mbox{if } \ux\in \RR\\ 
    \varphi_0\oplus \varphi_\AP\mapsto \ux(\varphi_\AP)  & \mbox{if } \ux\in \RB,
  \end{cases}
\end{equation}
for $\varphi_0\in C_0(\RR)$ and $\varphi_\AP\in \CAP$. Obviously, we have  
\begin{align}
\label{eq:iotaxiR}
	\xi(c) = \iota'(c)\qquad \forall\: c\in \RR,
\end{align} 
for $\iota'$ defined as $\iota$ in \eqref{eq:iota}, but now for $\varphi\in C_0(\RR)\oplus \CAP$. 

Moreover, \cite{InvConLQG} 
$\Borel(\qcR)=\Borel(\RR)\sqcup\Borel(\RB)$ holds for the Borel $\sigma$-algebra $\Borel(\qcR)$ of $\qcR$; and if $\mu\colon \Borel(\qcR)\rightarrow [0,\infty]$ is a Borel measure (i.e., locally finite; hence, finite by compactness of $\qcR$), then $\muR:=\mu|_{\Borel(\RR)}$ and $\muRB:=\mu|_{\Borel(\RB)}$ are (necessarily finite) Borel measures as well. This is because the subspace topologies of $\RR$ and $\RB$ w.r.t.\ the topology on $\qcR$ coincide with their usual ones. Then, $\muR$ and $\muRB$ are Radon if $\mu$ is Radon, and we have $\mu=\muR\oplus\muRB$ for
 \begin{align*}
	(\muR\oplus\muRB)(A):=\muR(A\cap \RR) + \muRB(A\cap \RB)\quad
\end{align*} 
for all $A\in \Borel(\qcR)$.
\section{Scaling Actions}
\label{sadasds}
Scale invariance is one of the conditions we will use to single out the measure on both LQC configuration spaces $\RB$ and $\RR\sqcup\RB$. For this, of course, we first need to lift the scaling action of $\RR_{\neq 0}$ on $\RR$ to these quantum spaces; which is the content of this section.

In \cite{InvConLQG} it was shown that, given a left action $\act\colon G\times X\rightarrow X$ and a unital $C^*$-subalgebra $\cC$ of the bounded functions on $X$ with 
\begin{align}
\label{eq:inv}
	\theta_g^*(\cC)\subseteq \cC\qquad\forall\:g\in G,
\end{align}
the left action
\begin{align*}
	\sact\colon G\times \Spec(\cC)\rightarrow \Spec(\cC),\quad (g,\x)\mapsto [\varphi\mapsto \x(\act_g^*\varphi)]
\end{align*}
is unique w.r.t.\ the property that for each $g\in G$,
\begingroup 
\setlength{\leftmargini}{12pt}
\begin{enumerate}
\item
	$\sact_g$ is continuous, and
\item
\itspace
	$\sact_g\cp \iota = \iota \cp \act_g$\: holds.
\end{enumerate}
\endgroup
\noindent
Here, the second condition means that
    \begin{center}
		\makebox[70pt]{
			\begin{xy}
				\xymatrix{
					\X\ar@{<-}[d]^-{\iota} \ar@{->}[r]^-{\Theta_g}  &  	\X\ar@{<-}[d]^-{{\iota}}   \\
					X\ar@{->}[r]^-{\theta_g} & X  
				}
			\end{xy}
		}
\end{center}
is commutative for each $g\in G$, i.e., that $\Theta$ extends $\theta$ in the canonical way.
 
Now, let $G=\RR_{\neq 0}$, $X=\RR$, $\cC=\CAP$ and
\begin{align*}
 \theta\colon \RR_{\neq 0} \times \RR \rightarrow \RR,\quad (t,c)\mapsto t\cdot c
\end{align*}
be the multiplicative action. Then, 
\begin{align*} 
 \theta_t^*\chi_\lambda=\chi_{t\cdot \lambda}\qquad \forall\: t\in \RR_{\neq 0},\quad\forall \:\lambda\in \RR, 
\end{align*} 
 hence $\theta_t^*(\CAP)\subseteq \CAP$ holds as $\theta_t^*$ is an isometry. Thus, $\theta$ extends uniquely to an action $\Theta\colon \RR_{\neq 0}\times \RB\rightarrow \RB$ that we will   
denote by $\Transl$ in the following. Then, we have 
\begin{align}
\label{jjsjsjs}
 \Transl(t,\iota(c))=\iota(t\cdot c)\qquad\forall\: t\in \RR_{\neq 0},\quad\forall\:c\in \RR
\end{align}  
by Condition 2., 
hence $\Transl_t\in \mathrm{Aut}(\RB)$ for each $t\in \RR_{\neq 0}$, by \eqref{eq:homomorph} and denseness of $\iota(\RR)$ in $\RB$, cf.\ also Corollary 4.2 in \cite{InvConLQG}.

Now, in order to provide an analogous $\RR_{\neq 0}$ action on $\qR$, we let 
\begin{align*}
  \Transwq(t,\ux) := 
  \begin{cases} 
    \hspace{10.1pt}t\cdot \ux &\mbox{if } \ux\in \RR\\ 
    \Transl(t, \ux) & \mbox{if } \ux\in \RB,
  \end{cases}
\end{align*}
and carry it over to $\qR$ via $\xi$; i.e., we define
\begin{align*}
\Transw(t,\x):=(\xi\cp \Transwq)\big(t,\xi^{-1}(\x)\big). 	
\end{align*}
Then, $\Transw$ equals the unique extension $\Theta$ of $\theta$ to $\qR$. 
In fact, first observe that $\Theta$ exists, because \eqref{eq:inv} obviously holds for $C_0(\RR)\oplus \CAP$ as well. Then, for $\x:=\xi(c)$ with $c\in \RR$, we have
\begin{align*}
	\Transw_t(\x)(\varphi)&=
	\xi(t\cdot c)(\varphi)=\varphi(t\cdot c)=(\theta_t^*\varphi)(c)\\[-2pt]
	&=\iota'(c)(\theta_t^*\varphi)
	\stackrel{\eqref{eq:iotaxiR}}{=}\xi(c)(\theta_t^*\varphi)=\x(\theta_t^*\varphi)\\[3pt]
	&=\Theta_t(\x)(\varphi)
\end{align*} 
for each $\varphi\in C_0(\RR)\oplus \CAP$, whereby the last step is due to the definition of $\Theta$. Moreover, for $\x:=\xi(\psi)$ with $\psi \in \RB$, we have
\begin{align*}
	\Transw_t(\x)(\varphi_0\oplus \varphi_\AP)&=\xi(\Transl(t,\psi))(\varphi_0\oplus \varphi_\AP)\\
	&=\xi(\Transl(t,\psi))(\varphi_\AP)
	=\psi(\theta_t^*\varphi_\AP)\\[1pt]
	&=\x(\theta_t^*\varphi_\AP)=\x(\theta_t^*(\varphi_0\oplus \varphi_\AP))\\
	&=\Theta_t(\x)(\varphi_0\oplus \varphi_\AP),
\end{align*} 
whereby, in the fifth step, we have used $\theta_t^*$-invariance of $C_0(\RR)$.
\section{Invariant Measures}
\label{sdsdsd}
Suppose that $\mu$ is a $\Transwq$-invariant Borel measure on $\RR\sqcup \RB\cong \qR$, i.e., that $\Transwq_t(\mu)=\mu$ holds for each $t\in \RR_{\neq 0}$, for $\Transwq_t(\mu)$ the push forward of $\mu$ by $\Transwq_t$. Then, $\mu=\muR\oplus \muRB$ holds for $\muR$ and $\muRB$ defined as in Sect.\ \ref{sec:prel}; whereby we must have
\begin{equation}
  \label{eq:Ksiiiaa}
  \muR(A) = 
  \begin{cases} 
    s &\mbox{if } 0\in A\\ 
    0  & \mbox{if } 0\notin A
  \end{cases}
\end{equation}
for some unique $s\geq 0$, and each $A\in \Borel(\RR)$. In fact, if $A$ is a bounded interval that contains $0$, invariance shows  $\mu(A)=s:=\mu(\{0\})$. This is clear from 
$\textstyle\bigcap_{n\in \NN}\Transl_{1/n}(A)=\{0\}$, 
because $\mu_\RR$ is continuous from above and finite. Then, since $\muR$ is monotonous, for each Borel set $A'$ with $0\in A'$ and $A'\subseteq A$, we have
\begin{align*}
	s=\muR(\{0\})\leq\muR(A')\leq \muR(A)=\muR(\{0\})=s,
\end{align*}
hence $\muR(A')=s$. In particular $\mu(A)=0$ holds if $A\in \Borel(\RR)$ is bounded with $0\notin A$, because $\mu$ is additive. Then, \eqref{eq:Ksiiiaa} is clear from $\sigma$-additivity of $\mu$. 
\vspace{6pt}

\noindent
Thus, it remains to investigate the $\Transl$-invariant normalized Radon measures $\muRB$ on $\RB$ in order to determine the scale invariant normalized Radon measures on both $\RR\sqcup \RB\cong \qR$ and $\RB$. Now, 
first it is clear from
\begin{align*}
	\Transl_t(\muB)(\psi \BP A)&\textstyle=\muB(\Transl(1/t,\psi)\BP \Transl(1/t,A))\\
	&=\muB(\Transl(1/t,A)) = \Transl_t(\muB)(A)
\end{align*}
for $\psi\in \RB$ and $A\in \Borel(\RB)$  
that the Haar measure $\muB$ on $\RB$ fulfils this requirement, just by uniqueness of the Haar measure up to scalation. For this, observe that $\Transl_t(\muB)$ is a normalized Radon measure on $\RB$ because  
$\Transl_t\colon \RB\rightarrow \RB$ is a homeomorphism (and an automorphism) for each $t\in \RR$. Second, in analogy to \eqref{eq:Ksiiiaa}, let us define
\begin{equation*}
  \mu_\delta(A) := 
  \begin{cases} 
    1 &\mbox{if } 0\in A\\ 
    0  & \mbox{if } 0\notin A,
  \end{cases}
\end{equation*}
but now for each $A\in \Borel(\RB)$. Obviously, $\mu_\delta$ is a $\Transl$-invariant normalized Radon measure on $\RB$; thus, the same is true for the measures
\begin{align*}
	\mu_z:=z\cdot \mu_\delta + (1-z)\cdot \muB\quad\forall\: 0\leq z\leq 1. 
\end{align*} 
We now show that these are the only $\Transl$-invariant normalized Radon measures on $\RB$. For this, let $\muRB$ be such a measure; and observe that, by the Riesz-Markov theorem\footnote{In the following, we refer to the Riesz-Markov theorem in the form of 2.5 Satz in Kapitel VIII. in \cite{Elstrodt}.}, it is uniquely determined by the positive continuous ($\muRB$ was assumed to be normalized) linear functional
\begin{align*}
	\textstyle L\colon C(\RB)\rightarrow \CC,\quad \alpha\mapsto \int \alpha \:\dd\muRB.
\end{align*} 
Since the characters $\{\chi_\lambda\}_{\lambda\in \RR}$ span a dense subset $\VD\subseteq \CAP$, their Gel'fand transforms $\{\gelhat{\chi}_\lambda\}_{\lambda\in \RR}$ span a dense subset of $C(\RB)$. Thus, by continuity and linearity of $L$, $\muRB$ is uniquely determined by the values $L(\gelhat{\chi}_\lambda)$, whereby
\begin{align*}
\textstyle L(\gelhat{\chi}_0)=\int 1\:\dd\muRB=\muRB(\RB)=1
\end{align*}
holds. 
Then, invariance shows  
\begin{align*}
	\textstyle L(\gelhat{\chi}_\lambda)&\textstyle= \int \gelhat{\chi}_\lambda\:\dd\muRB=\int \gelhat{\chi}_\lambda\:\dd\Transl_{t}(\muRB)\\
	&\textstyle=\int \gelhat{\chi}_\lambda\cp\Transl_t\:\dd\muRB=\int \gelhat{\chi}_{t\cdot \lambda}\:\dd\muRB=L(\gelhat{\chi}_{t\cdot \lambda})
\end{align*}
for each $\lambda, t\neq 0$; hence 
\begin{align*}
	\textstyle L(\gelhat{\chi}_\lambda)=z=z\cdot \gelhat{\chi}_\lambda(0)=z\cdot \int \gelhat{\chi}_\lambda\:\dd\mu_\delta\quad\forall\: \lambda\neq 0
\end{align*}
for some $z\in \CC$. Here, we must have $-1\leq z\leq 1$, because 
\begin{align*}
	\textstyle|z|=|L(\gelhat{\chi}_\lambda)|\leq \int |\gelhat{\chi}_\lambda|\: \dd\muRB=1 
\end{align*}	
	and $z=L(\gelhat{\chi}_{-\lambda})=\ovl{L(\gelhat{\chi}_\lambda)}=\ovl{z}$ holds. 
Then, since we have $\int \gelhat{\chi}_\lambda \:\dd\muB=0$ for each $\lambda\neq 0$ by general theory of locally compact abelian groups \footnote{This is because $\RB$ is compact, and since $\gelhat{\chi}_\lambda\neq 1$ is a continuous character on $\RB$, cf.\ \cite{RudinFourier}.},  
\begin{align*} 
	\textstyle L(\gelhat{\chi}_\lambda)&\textstyle=z\cdot \int \gelhat{\chi}_\lambda \:\dd\mu_\delta + (1-z)\cdot \int \gelhat{\chi}_\lambda\:\dd\muB
\end{align*}
holds for each $\lambda\in \RR$; hence,
\begin{align}
\label{oapaspoaposa} 
	\textstyle L(\alpha)=
	z\cdot \int \alpha \:\dd\mu_\delta + (1-z)\cdot \int \alpha\:\dd\muB 
\end{align}
for each $\alpha\in C(\RB)$ by continuity of the linear forms representing both integrals, as well as by denseness of $\GEL(\VD)$ in $C(\RB)$. 
 Now, the Riesz-Markov correspondence \cite{Elstrodt} shows  
\begin{align*}
	\muRB(\{0\})&=\inf\{\hspace{2.8pt}L(\alpha)\:|\: C(\RB)\ni\alpha\geq 0: \alpha(0)=1 \}\\
	\muB(\{0\})&=\inf\{L'(\alpha)\:|\: C(\RB)\ni\alpha\geq 0: \alpha(0)=1  \},
\end{align*}
for $L'$ 
the linear form representing $\muB$. Then, \eqref{oapaspoaposa} shows that $0\leq \muRB(\{0\})= z$ holds, because we have $\muB(\{0\})=0$. Thus, we have $\muRB=\mu_z$ for some $0\leq z\leq 1$ by \eqref{oapaspoaposa}. 

\section{Uniqueness}
\label{sdsdsdsd}
In the previous section, we have determined the scale invariant normalized Radon measures on $\RB$ and $\RR\sqcup \RB$. In this section, we are going to single out $\muB$ on both configuration spaces by the requirement that there exist 
a momentum operator which both satisfies the correct commutation relations and is hermitian.  
This is a condition which arises by considering the next step in the quantization, namely the representation of basic operators.
\subsection*{QUANTIZATION}
The basic variables of each quantization scheme are the elements of the configuration algebra $\cC$ selected
($\CAP$ in the standard case, and $C_0(\RR)\oplus \CAP$ in the Fleischhack one), together with the momentum $p$.
The classical Poisson algebra satsified by these variables is 
\begin{align}
\label{clpoisson}
\{p, \varphi\} = - \dot\varphi 
\end{align}
which has to be understood as an equation on the dense subalgebra $\DIDE\subseteq \cC$ of all $\cC\ni \varphi\colon \RR\rightarrow \CC$ that are of class $C^\infty$ with $n$th derivative  contained in $\cC$ for each $n\in\NN$. 

Here, denseness is clear for the standard case where $\cC=\CAP$ holds, since there we have $\VD\subseteq \DIDE$ for $\VD$ the span of the $\{\chi_\lambda\}_{\lambda\in \RR}$. In the Fleischhack case where $\cC=C_0(\RR)\oplus \CAP$ holds, we find that 
\begin{align}
\label{ges}
C_0(\RR)\ni \varphi_\epsilon\colon \RR\rightarrow \RR\subseteq \CC,\quad c\mapsto \mathrm{e}^{-(\epsilon+c)^2}
\end{align}
is contained in $\DIDE$ for each $\epsilon \in \RR$; whereby the products $\{\varphi_0\cdot \chi_\lambda\}_{\lambda\in \RR}\subseteq \DIDE$ separate the points in $\RR$. Thus, by Stone-Weierstrass, they generate a dense $^*$-subalgebra of $C_0(\RR)$; showing that $\DIDE$ is dense in $\cC$ as it also contains $\VD$. 

Now, the states of the system are wavefunctions  
on the quantum configuration space -- i.e., the Gel'fand spectrum $\Spec(\cC)$ of the configuration algebra. More precisely, such a wavefunction is an element of the Hilbert space $\Hil:=L^2(\Spec(\cC),\mu)$ of all $f\colon \Spec(\cC)\rightarrow \CC$ that are square integrable w.r.t.\ the normalized Radon measure $\mu$ to be determined in the following.     
Here, we can consider the subspace $\mathcal{G}(\DIDE)$ arising as the Gel'fand transform of elements of the algebra $\DIDE$, a subspace which is dense in $\Hil$ because $\GEL(\DIDE)$ is dense in $C(\Spec(\cC))=\GEL(\cC)$ by definition, and since $\GEL(\cC)$ is dense in $\Hil$ by the properties of $\mu$.   
\footnote{We have $L^2(\Spec(\cC),\mu)\subseteq L^1(\Spec(\cC),\mu)$ because $\mu$ is finite (cf.\ 2.10 Satz in Kapitel VI. in \cite{Elstrodt}). Moreover, $C(\Spec(\cC))=\GEL(\cC)$ is dense in $L^1(\qR,\mu)$ because $\mu$ is Radon (combine 2.28 Satz in Kapitel VI. in \cite{Elstrodt} with Riesz-Markov).}
On this Hilbert space, the operators $\widehat{\varphi}$ corresponding to each $\varphi\in \cC$ act by multiplication, $\widehat{\varphi}\colon \Hil\ni f\mapsto \GEL(\varphi) \cdot f$, 
while the momentum operator $\widehat{p}$ must be chosen to satisfy commutation relations mimicking (\ref{clpoisson}):
\begin{align}
\label{basiccomm}
\textstyle[\widehat{p}, \widehat{\varphi}] = - \I \cdot \widehat{\dot\varphi}\qquad\forall\:\varphi\in \DIDE,
\end{align}
now, to be understood as an equation of operators defined on the dense subset $\GEL(\DIDE) \subseteq \Hil$. Applying both sides to 
$\gelhat{\chi}_0\equiv 1$, we find that
\begin{align}
\label{genp}
\textstyle\widehat{p} = -\I\cdot \mathcal{G} \circ \frac{\dd}{\dd c} \circ \mathcal{G}^{-1} +\I\cdot f
\end{align}
holds on $\GEL(\DIDE)$ for $f:=-\I \cdot \widehat{p}(\gelhat{\chi}_0)\in \Hil$. 
This is the most general form of momentum operator defined on $\GEL(\DIDE)$ satisfying the basic commutation relations.

The final requirement that $f$ can be chosen so that $\widehat{p}$ is hermitian on a suitable dense subset of $\cC$ will impose a restriction 
on the inner product and hence an additional restriction on the measure.
We derive this restriction separately for the standard and the Fleischhack case. 

Since hermicity of $\widehat{p}$ does not depend on the imaginary part of $f$ in \eqref{genp}, we can assume without
loss of generality that $\mathrm{im}[f]\subseteq \RR\sqcup\{-\infty,\infty\}$ holds in the following. 
With this assumption we will find that hermicity of $\widehat{p}$ enforces $f=0$.
Thus, without this assumption, the conclusion is that $f$ must be pure imaginary, so that each momentum operator fulfilling \eqref{basiccomm} on $\GEL(\DIDE)$, can be written in the form 
\begin{align}
\label{hjhjhjg}
\textstyle\widehat{p} = -\I\cdot \mathcal{G} \circ \frac{\dd}{\dd c} \circ \mathcal{G}^{-1} +g
\end{align}
for some $g\in \Hil$ with $\mathrm{im}[g]\subseteq \RR\sqcup \{-\infty,\infty\}$.  
Furthermore, with an additional minor assumption,
even this remaining ambiguity in the quantization of $\widehat{p}$ is not physical, as the resulting quantizations 
are unitarily equivalent, as reviewed in the appendix.

\subsection*{THE STANDARD CONFIGURATION SPACE} 
\label{Stan}
We already know that each scale invariant normalized Radon measure on $\RB$ is of the form 
\begin{align*}
\muRB=\mu_z=z\cdot \mu_\delta + (1-z)\cdot \muB
\end{align*}
for some unique $0\leq z\leq 1$. 
We denote the corresponding $L^2$-Hilbert space by $\Hil$, fix some $f\in \Hil\subseteq L^2(\RB,\muB)$ 
with $\mathrm{im}[f]\subseteq \RR\sqcup\{-\infty,\infty\}$,  
and define $\wpo$ by \eqref{genp}; hence 
\begin{align*}
	\wpo(\gelhat{\chi}_\lambda)=(\lambda+\I\cdot f)\cdot \gelhat{\chi}_\lambda\quad\forall\: \lambda\in \RR.
\end{align*}	 
We now require $\wpo$ to be hermitian on $\VD$, i.e., that 
\begin{align}
\label{apodspsdaoa}
	\langle \wpo (\gelhat{\chi}_\lambda)|\gelhat{\chi}_{\lambda'}\rangle=\langle \gelhat{\chi}_\lambda|\wpo (\gelhat{\chi}_{\lambda'})\rangle\quad\forall\:\lambda\in \RR
\end{align}
holds. This means
\begin{align*}
	0&=\langle \wpo (\gelhat{\chi}_\lambda)|\gelhat{\chi}_{\lambda'}\rangle -\langle \gelhat{\chi}_\lambda|\wpo (\gelhat{\chi}_{\lambda'})\rangle\\
	&\textstyle=(\lambda-\lambda')\cdot \langle \gelhat{\chi}_\lambda|\gelhat{\chi}_{\lambda'}\rangle - 2\I\cdot \langle \gelhat{\chi}_\lambda|f\cdot \gelhat{\chi}_{\lambda'}\rangle\\
	&= z\cdot(\lambda-\lambda') - 2\I\cdot \langle \gelhat{\chi}_{\lambda-\lambda'}| f\rangle 
\end{align*}
for all $\lambda,\lambda'\in \RR$; so that \eqref{apodspsdaoa} holds iff 
	$0=z \lambda - 2\I\cdot \langle \gelhat{\chi}_{\lambda}| f\rangle$ holds for each $\lambda\in \RR$, 
hence 
\begin{align*}
	\textstyle\frac{z \lambda}{2\I}=z\cdot f(0)+(1-z)\cdot\int \gelhat{\chi}_{-\lambda}\cdot f\:\dd\muB \quad\forall\:\lambda\in \RR.
\end{align*}
In particular, we must have $0\leq z<1$, hence
\begin{align*}
	\textstyle\int \gelhat{\chi}_{-\lambda}\cdot f\:\dd\muB=\textstyle\frac{z }{1-z}\cdot\big(\frac{\lambda}{2\I}-f(0)\big)\quad\forall\:\lambda\in \RR.
\end{align*}
For $z>0$, this is non-zero for each $\lambda\neq 0$, so that we cannot have $f\in \Hil\subseteq L^2(\RB,\muB)$ since $\RR_{\neq 0}$ is uncountable. Thus, we must have $z=0$, hence $\muRB=\muB$ and $f=0$.

\subsection*{THE FLEISCHHACK CONFIGURATION SPACE}
\label{Fan}
As shown in Sect.\ \ref{sdsdsd}, the scale invariant normalized Radon measures on $\RR\sqcup \RB\cong\qR$ are of the form 
\begin{align*}
\mu=\mu_{u,z}=u\cdot \mu_\RR \oplus (1-u)\cdot \mu_z
\end{align*}
 for some $0\leq u\leq 1$, some $0\leq z\leq 1$, and $\mu_\RR$ defined by \eqref{eq:Ksiiiaa} for $s=1$. 
We let $\wpo$ be defined by \eqref{genp} for $f\in L^2(\qR,\xi(\mu))$ with $\mathrm{im}[f]\subseteq \RR\sqcup\{-\infty,\infty\}$, and observe that hermicity of $\widehat{p}$ (now, supposed to hold on the dense subalgebra generated by the function $\varphi_\epsilon$ and the elements in $\VD$) enforces
\begin{align*}
	0&=\langle \wpo (\GEL(\varphi_\epsilon))|\gelhat{\chi}_{0}\rangle -\langle \GEL(\varphi_\epsilon)|\wpo (\gelhat{\chi}_{0})\rangle\\
	&=\hspace{3pt}\I\cdot \langle \GEL(\dot{\varphi}_\epsilon)|\gelhat{\chi}_{0}\rangle-2\I\cdot \langle \GEL(\varphi_\epsilon)| f\rangle,
\end{align*}
for $\varphi_\epsilon$ defined by \eqref{ges} for each $\epsilon \in \RR$; hence,
\begin{align}
\label{ooopquztfd}
\begin{split}
	 0&\textstyle=\int (\GEL(\dot{\varphi}_\epsilon)-2 f\cdot \GEL(\varphi_\epsilon))\: \dd\xi(\mu)\\
	&\textstyle= \int(\dot{\varphi}_\epsilon-(2 f\cp\iota')\cdot \varphi_\epsilon)\: \dd(u\cdot \mu_\RR)\\
	&\textstyle=u\cdot(\dot{\varphi}_\epsilon(0) - (2 f\cp\iota')(0)\cdot \varphi_\epsilon(0)).
	\end{split}
\end{align}
The second step is due to the fact that
\vspace{-4pt}
\begin{align*}
	\GEL(h)|_{\xi(\RB)}=0\qquad\text{and}\qquad h\cdot \cC\subseteq C_0(\RR)
\end{align*}
holds for each $h\in C_0(\RR)$, where the first equation is from \eqref{eq:Ksiii}.
To see how this implies the second step, first observe that $L^2(\qR,\mu)\subseteq L^1(\qR,\mu)$ holds, because $\mu$ is finite; and that $C(\qR)=\GEL(\cC)$ is dense in $L^1(\qR,\mu)$, because $\mu$ is Radon. 
So, we can choose a sequence $C(\qR)\supseteq \{f_n\}_{n\in \NN}\rightarrow f$ converging w.r.t.\ the $L^1$-norm, and obtain 
\begin{align*}
	\textstyle\int f\cdot \:&\GEL(\varphi_\epsilon)\:\dd\xi(\mu)\textstyle=\lim_n \int f_n\cdot \GEL(\varphi_\epsilon)\:\dd\xi(\mu)\\
	&\textstyle=\lim_n \int ((f_n\cdot \GEL(\varphi_\epsilon))\cp\xi)\:\dd\mu\\
	&\textstyle=\lim_n \int_\RR ((f_n\cp\iota')\cdot \varphi_\epsilon)\:\dd(u\cdot \mu_\RR)\\
	 &\quad\:\textstyle + \lim_n \int_{\RB} ((f_n\cdot \GEL(\varphi_\epsilon))\cp \xi)\:\dd((1-u)\cdot \mu_z),
\end{align*}
whereby the second integral is zero as $f_n\cdot \GEL(\varphi_\epsilon)\in \GEL(C_0(\RR))$ holds. 
\vspace{4pt}

\noindent
Now, \eqref{ooopquztfd} is obviously fulfilled for $u=0$, in which case we have $\mu=\mu_z$. 
Suppose by way of contradiction that the other case holds. 
By choosing $\epsilon=0$ we obtain in this case $(f\cp\iota')(0)=0$. Then, choosing $\epsilon \neq 0$, we conclude that $u=0$ holds, which contradicts $u>0$.  

Thus, we must have $\mu=\mu_z$ for some $0\leq z\leq 1$; so that
\begin{align*}
\rho\colon \Hil \rightarrow L^2(\RB,\mu_z), \quad\alpha\mapsto (\alpha\cp \xi)|_{\RB}
\end{align*}
is unitary by
\begin{align*}
\textstyle(\|\alpha\|_2)^2=\int |\alpha\cp\xi|^2\: \dd\mu_z= \int |\rho(\alpha)|^2\: \dd\mu_z.
\end{align*} 
Then, hermicity of $\wpo$ reduces to condition \eqref{apodspsdaoa}; so that the same arguments as in 
the prior subsection show $z=0$ and $\rho(f)=0$.

\newpage

\begin{acknowledgments}
This work has been supported in part by the Alexander von Humboldt foundation of Germany and 
NSF Grant PHY-1505490.
\end{acknowledgments}

\appendix
\section*{APPENDIX: Unitary Equivalence}
As we have seen, hermicity enforces $\widehat{p}$ to be of the form \eqref{hjhjhjg} for some $g=\widehat{p}(\gelhat{\chi}_0)\in \Hil$ with $\mathrm{im}[g]\subseteq \RR\sqcup\{-\infty,\infty\}$.    
Then, if we additionally require that, besides \eqref{basiccomm}, also the relation $[\widehat{p}, \widehat{p}] = 0$ makes sense, we must have $\widehat{p}\colon \GEL(\DIDE)\rightarrow\GEL(\DIDE)$, hence $g\in C(\Spec(\cC))$ with $\mathrm{im}[g]\subseteq \RR$. 
Since $\cC$ consists of continuous (and bounded) functions on $\RR$, the real valued function $h:=\GEL^{-1}(g)$ admits an antiderivative $H\colon \RR\rightarrow \RR$. 

We now  make the further assumption that $H$ is an element of $\cC$, i.e., that the unitary operator
\begin{align*}
	U\colon \Hil\rightarrow \Hil,\quad \alpha\mapsto \alpha\cdot \mathrm{e}^{-\I\cdot \GEL(H)}
\end{align*}
is well defined. Obviously, then we have  $U^* \cp \widehat{\varphi} \cp U= \widehat{\varphi}$ for each $\varphi\in \DIDE$; and $U$ also intertwines $\widehat{p}$ and $\widehat{p}_0:=-\I\cdot \mathcal{G} \circ \frac{\dd}{\dd c} \circ \mathcal{G}^{-1}$  
as one can check.  For this, one needs to use 
$\mathrm{e}^{-\I \cdot\GEL(H)}=\GEL\cp\mathrm{e}^{-\I \cdot H}$, 
which follows from the fact that $\GEL$ is an isometry and an algebra isomorphism, and that  
\begin{align*}
	\textstyle\mathrm{e}^{-\I \cdot H}=\sum_{n=0}^\infty \frac{1}{n!} (-\I \cdot H)^n
\end{align*} 
converges uniformly to some element in $\cC$ as we have $H\in \cC$.

\end{document}